\documentclass[aps,prb,twocolumn,footinbib,groupedaddress,longbibliography]{revtex4-1}
\usepackage{graphicx}
\usepackage{subfigure}
\usepackage{dcolumn}
\usepackage{bm}
\usepackage{times}
\usepackage{amsmath}
\usepackage{amssymb}
\usepackage{float}
\usepackage{color}
\usepackage{multirow}
\usepackage{url,hyperref}
\usepackage[normalem]{ulem}

\begin{document}
\title{Current-induced near-field radiative energy, linear-momentum, and angular-momentum
transfer}
\author{Huimin Zhu,$^{1,4}$ Gaomin Tang,$^{2,*}$ Lei Zhang,$^{1,4,\dagger}$ and Jun
Chen$^{3,4,\ddag}$}
\affiliation{$^1$State Key Laboratory of Quantum Optics and Quantum Optics Devices, Institute
  of Laser Spectroscopy, Shanxi University, Taiyuan 030006, China\\
  $^2$Graduate School of China Academy of Engineering Physics, Beijing 100193, China\\
  $^3$State Key Laboratory of Quantum Optics and Quantum Optics Devices, Institute of
  Theoretical Physics, Shanxi University, Taiyuan 030006, China\\
  $^4$Collaborative Innovation Center of Extreme Optics, Shanxi University, Taiyuan
  030006, China}

\bigskip

\begin{abstract}
In this paper, we study the near-field radiative energy, linear-momentum, and angular-momentum transfer from a current-biased graphene to nanoparticles. The electric current through the graphene sheet induces nonequilibrium fluctuations, causing energy and momentum transfer even in the absence of a temperature difference. The inherent spin-momentum locking of graphene surface plasmons leads to an in-plane torque perpendicular to the direction of the electric current. In the presence of a temperature difference, the energy transfer is greatly enhanced while the lateral force and torque remain within the same order. Our work explores the potential of utilizing current-biased graphene to manipulate nanoparticles.
\end{abstract}

\maketitle

\section{Introduction}
The inclusion of nonreciprocal effects in the study of near-field thermal radiation has
led to the discovery of various novel phenomena and has attracted significant scientific
interest~\cite{control_magnetic_15, nonrecip_Hall, nonrecip_persistent,
control_magnetic_17, control_magnetic_18, nonrecip_rect, GT_WSM, GT21,review21}.
Nonreciprocity can be achieved in various scenarios. One example is by applying a magnetic
field to magneto-optic materials~\cite{control_magnetic_15, nonrecip_Hall,
nonrecip_persistent, control_magnetic_17, control_magnetic_18, nonrecip_rect}.
Another approach is to use magnetic Weyl semimetals that have intrinsic time-reversal
symmetry breaking~\cite{WSM_radiate1, WSM_radiate2, WSM_radiate3, WSM_radiate4}.

Nonreciprocity-induced lateral force has been studied recently~\cite{lateral_17, lateral_21, lateral_21-2, lateral_23}.
An atom or nanoparticle placed above a sample with mobile carrier drift experiences a fluctuation force~\cite{lateral_17}.
In a system where vacuum-separated plates are at different temperatures, if at least one plate is nonreciprocal material, the setup can function as a heat engine where thermal energy is converted into mechanical work~\cite{lateral_21}.
When a dipolar particle is placed in the vicinity of nonreciprocal plasmonic slab, it experiences a lateral force and torque in the presence of a temperature difference between the particle and the slab~\cite{lateral_21-2}. The force is linked to the nonreciprocity of the surface polaritons and the torque to the spin-momentum locking of the surface polaritons.

Current-biased graphene is also known for its nonreciprocity~\cite{drift_15, drift_16, drift_18, drift_20}.
The exceptional electron mobility of graphene allows it to sustain nonreciprocal surface plasmon  polaritons (SPPs)
when driven by an electric current, which gives rise to interesting properties such as Fizeau drag~\cite{Fizeau1, Fizeau2}
and negative Landau damping~\cite{drift_17}. Current-biased graphene is in a nonequilibrium state and this leads to
a finite photonic chemical potential~\cite{photon_noise_RMP} that depends on the in-plane wave vector for the thermal
electromagnetic radiation. The occupation number of radiative photons is nonreciprocal, that is asymmetric with respect to
the positive and negative in-plane wave vectors~\cite{VP08_PRB, VP11_PRL, VP11_PRB, drift1}.
The nonreciprocal properties of current-biased graphene can be adjusted through chemical doping or gate voltage,
making it suitable for a wide range of applications in controlling thermal radiation. For example,
near-field radiative heat transfer has been studied between two suspended graphene sheets~\cite{drift2},
graphene covered substrates~\cite{shi2022near}, graphene-nanoparticles~\cite{nonrecip_diode},
and graphene-based multilayer systems~\cite{zhou2020enhancement}. However, the physical aspects of
the force and torque experienced by the nanoparticle near the current-biased graphene remain unexplored to date.

In this paper, we investigate the near-field transfer of energy, linear momentum, and angular
momentum between an isotropic dipolar nanoparticle and the current-biased graphene.
The nonequilibrium state in the current-biased graphene induces energy transfer between the nanoparticle
and the graphene even in the absence of a temperature difference. Importantly, these SPPs carry both linear and spin-locked angular momentum. As a result, there is an exchange of net linear and spin angular momentum
between the particle and the SPPs, leading to both lateral force and torque experienced by the nanoparticle.

\begin{figure*}
\centering
\includegraphics[width=\textwidth]{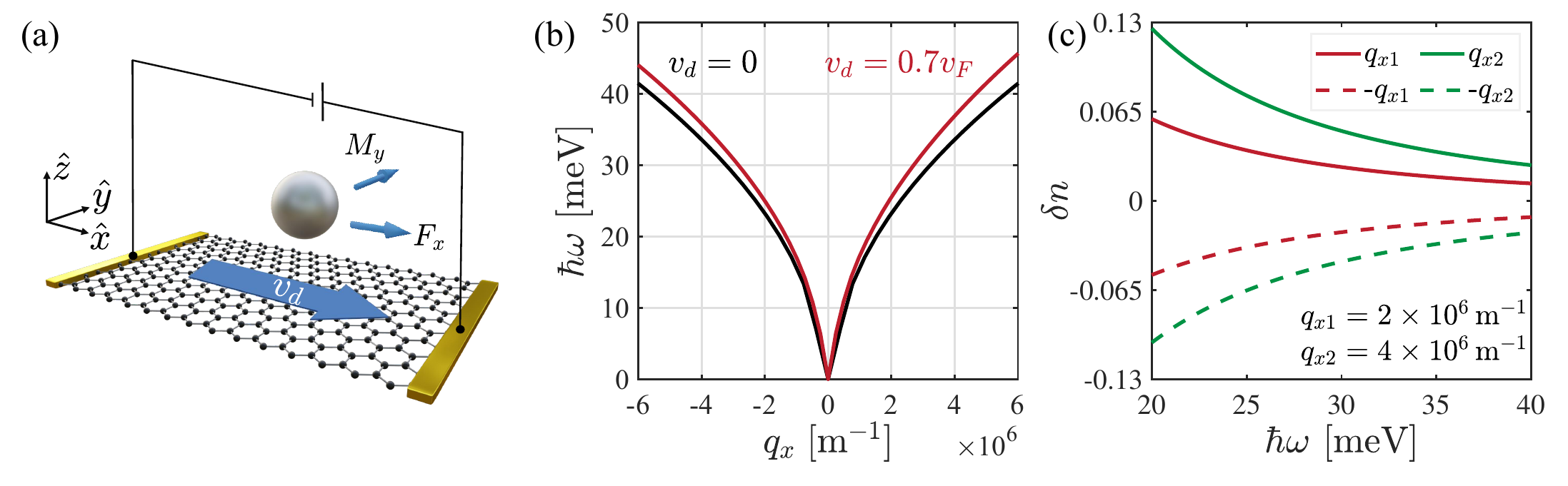} \\
\caption{(a) Schematic plot of the lateral force and torque exerted on an
  isotropic dipolar particle above a current-biased graphene in the near-field regime. The drift
  velocity $v_d$ is along the $x$ direction. (b) Dispersion of the SPPs
  of the current-biased graphene. The electric current
  effectively drags the SPPs which leads to the property of nonreciprocal propagation.
  (c) The photon occupation number difference $\delta n(\omega, q_x)$
  vs $\omega$ with $v_d=0.7v_F$. }
\label{fig1}
\end{figure*}

\section{System and formalism}
The system is illustrated schematically in Fig.~\ref{fig1}(a), where the spherical
nanoparticle with radius $R$ is placed near the surface of current-biased monolayer graphene.
Without losing generality, we assume that the drift velocity $v_d$ in graphene is along
the positive direction of the $x$ axis.
Within the framework of fluctuational electrodynamics, the total power $H$, lateral
force along the $x$ direction $F_x$, and torque along the $y$ direction $M_y$ can be expressed as
\begin{equation}
 P = \int_0^{\infty} \frac{d\omega}{2\pi} p(\omega),
\end{equation}
with $P\in \{H, F_x, M_y\}$ and the corresponding spectral densities $p\in \{h,f_x,m_y\}$.
The spectral densities are given by~\cite{lateral_21-2}
\begin{align}
\label{EqP}
  h(\omega) &= -\hbar\omega k_0^2 {\rm Im}[\alpha(\omega)] \int\frac{d^2{\bm q}}{(2\pi)^2}
  \Theta(\omega, {\bm q}) \delta n(\omega, q_x) , \\
  f_x(\omega) &= \hbar k_0^2 {\rm Im}[\alpha(\omega)] \int\frac{d^2{\bm q}}{(2\pi)^2} q_x
  \Theta(\omega, {\bm q}) \delta n(\omega, q_x) , \\
  m_y(\omega) &= -\hbar {\rm Im}[\alpha(\omega)] \int\frac{d^2{\bm q}}{(2\pi)^2} q_x
  {\rm Im}\big( r_{p} e^{2i\beta_0 d} \big) \delta n(\omega, q_x) ,
\end{align}
with
\begin{equation}
  \Theta(\omega, {\bm q}) = {\rm Re}\bigg[\frac{r_p e^{2i\beta_0 d}}{2\beta_0}
  \left(\frac{2q^2}{k_0^2} -1\right) \bigg] ,
\end{equation}
where the Fresnel reflection coefficient for $p$ polarization is given by
\begin{equation}
  r_p = \frac{\beta_0 \sigma_g  }{2 \epsilon_{0} \omega + \beta_0\sigma_g }.
\end{equation}
Here, $\sigma_g$ is the surface conductivity of graphene, and $\epsilon_{0}$ is the dielectric constant in vacuum. The in-plane wave
vector and the angular frequency are denoted by ${\bm q}=(q_x, q_y)$ and $\omega$.
The magnitude of the out-of-plane wave vector in air is $\beta_0 = \sqrt{k_0^2 -q^2}$ with
$k_0=\omega/c$ and $q=|{\bm q}|$.
When an electric current is applied to the graphene so that the electrons near the Fermi surface experience a drift velocity, the system is not in local thermal equilibrium and the occupation number of the
photons radiated from graphene is Doppler shifted. By assuming a drift velocity $v_d$ in the $x$ direction, the photon-occupation difference
between the graphene and the nanoparticle is~\cite{VP08_PRB,VP11_PRL,VP11_PRB, drift1, GT21}
\begin{equation} \label{delta_n}
  \delta n(\omega, q_x) = [e^{\hbar(\omega - q_x v_d)/k_B T_e} -1]^{-1} -
  [e^{\hbar\omega/k_B T_p} -1]^{-1} ,
\end{equation}
where $T_e$ and $T_p$ are the temperatures of graphene and the nanoparticle,
respectively.
In getting the occupation number of the photons radiated from the current-biased graphene, one needs to apply the condition that $\hbar q_x v_d$ is much smaller than the chemical potential~\cite{drift1}.
Equation~\eqref{delta_n} indicates that the electric current-induced
nonequilibrium fluctuations can produce a finite energy flux even in the absence of a
temperature difference. In the following, we assume that the
temperature of the particle and the graphene is the same with $T_p=T_e=T=300\,$K, unless there is a special
account for the temperatures. The energy transmission function of the energy flux in Eq.~\eqref{EqP} is defined through
$h(\omega)=\int d^2{\bm q} {\cal Z}_h(\omega, {\bm q})/4\pi^2$ with
\begin{equation}
  {\cal Z}_h(\omega, {\bm q}) = -\hbar\omega k_0^2 {\rm Im}[\alpha(\omega)]\Theta(\omega, {\bm q}) \delta n(\omega, q_x),
\end{equation}
which provides the information of the energy transfer at given $\omega$ and $\bm q$. Similar definitions are, respectively, applied to the force and torque.

\begin{figure*}
\centering
\includegraphics[width=\textwidth]{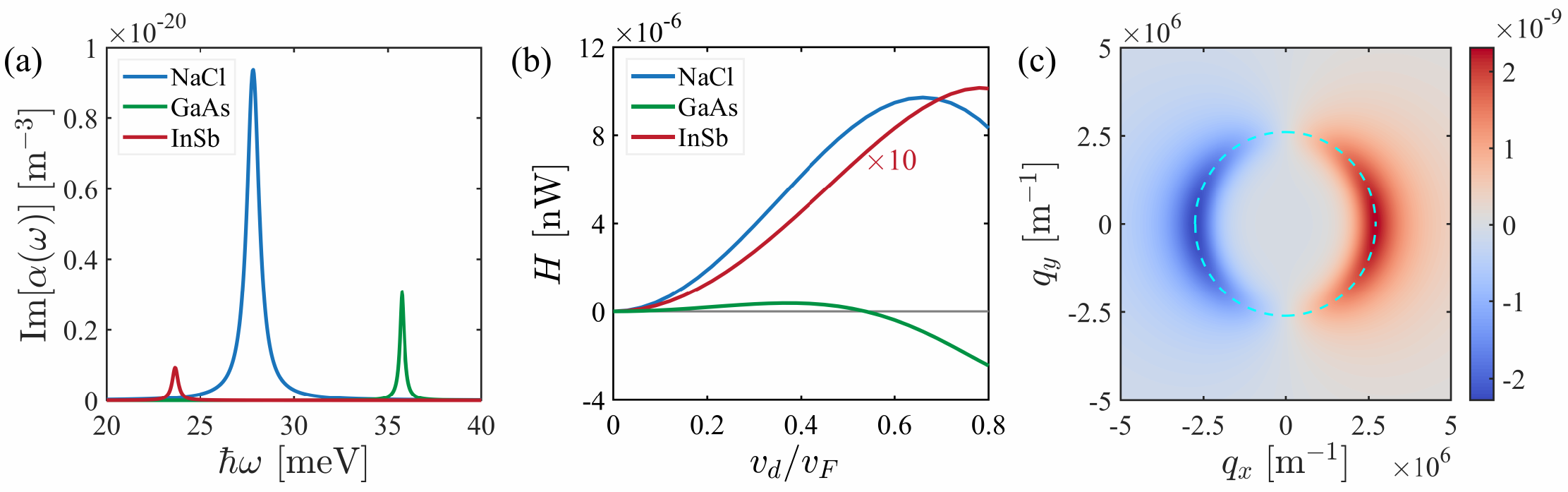} \\
\caption{(a) Imaginary part of the polarizability for different nanoparticles with radius $R = 40\,$nm. (b) Net power transfer experienced by nanoparticles as a function of $v_d$ at a separation distance of $d = 0.2\,\mu$m. The curve for the InSb particle is multiplied by a factor of $10$ for better visibility. (c) Energy transmission function ${\cal Z}_h$ (in units of nJ$\,$nm$^2$) plotted against $q_x$ and $q_y$ at $v_d = 0.3v_F$, $d = 0.2\,\mu$m, and $\hbar\omega =27.8\,$meV. The cyan dashed line represents the dispersion of graphene SPPs.}
\label{fig2}
\end{figure*}

The dispersion relation of SPPs in a monolayer graphene can be obtained through~\cite{graphene_plasmon}
\begin{equation}
  2\varepsilon_0 \omega + \beta_0\sigma_g =0,
\end{equation}
where the nonlocal surface conductivity $\sigma_g$ is given by~\cite{ drift_17_c1}
\begin{align}
  \sigma_g(\omega,q_x,q_y) &= \frac{ie_0^2\omega}{q^2} \frac{\mu(T)}{(\pi\hbar v_F)^2} \int_0^{2\pi} d\theta \frac{1}{(1-\cos\theta v_d/v_F)^2} \times \notag \\
  & \frac{q_x(\cos\theta-v_d/v_F) + q_y\sin\theta}{(\hbar\omega+i\gamma) /
  (\hbar v_F)-q_x \cos\theta -q_y\sin\theta} ,
\end{align}
with $\mu(T)=2k_BT\ln [2\cosh(\mu_g/2k_BT)]$. Here, $T$ is the temperature, $\mu_g$ is the chemical potential, $\gamma$ the damping parameter, and $e_0$ the electron charge.
The nonlocal surface conductivity is obtained under the long-wavelength limit and by neglecting the interband transition
which is justifiable for $2\mu_g > \hbar\omega$.
In the numerical calculation, we set the Fermi velocity $v_F=10^6\,$m/s,
the chemical potential $\mu_g = 0.1\,$eV and
the damping parameter $\gamma = 3.7\,$meV. Figure~\ref{fig1}(b) shows the nonreciprocal
dispersions of SPPs in the presence of an electric current.

\section{Numerical Results}
Before considering the nonequilibrium energy and momentum transfer, we
discuss the nonreciprocal properties of $\delta n(\omega, q_x)$ shown in
Fig.~\ref{fig1}(c). We can see that the photon occupation number difference $\delta
n(\omega, q_x)$ is positive for positive $q_x$ under $\omega {>}q_xv_d$. This indicates that
the effective photonic temperature of the graphene is higher than the
particle temperature $T_p$. Furthermore, it should be noted that contributions from
$\omega {<} q_xv_d$ can be disregarded. This is because these contributions become
significant only when the separation between the particle and the graphene reaches
subnanometer scales~\cite{VP11_PRB}, whereas in our case, the distance is on the
order of micrometers. Consequently, photons, especially those with positive $q_x$, flow
from the graphene to the particle. In contrast, for negative $q_x$, $\delta n(\omega, q_x)$ is always negative. Therefore, the effective photonic temperature of the graphene is lower than $T_p$. This sign reversal indicates a reversal in the direction of photon transfer compared to the behavior exhibited by positive $q_x$.

We consider three kinds of nanoparticle: NaCl, undoped GaAs, and InSb.
The intricate dielectric characteristics are given by the Lorentz-Drude model with
\begin{equation}
  \varepsilon(\omega) = \varepsilon_{\infty} \frac{\omega_L^2-\omega^2-i\Gamma
  \omega}{\omega_T^2-\omega^2-i\Gamma\omega},
\end{equation}
where $\varepsilon_\infty$ is the high-frequency dielectric constant, and $\omega_L$ is the
longitudinal optic-phonon frequency, $\omega_T$ is the transverse optic-phonon frequency,
$\Gamma$ is the optic-phonon damping constant. Within the electrostatic limit, we can reasonably neglect multiple scattering interactions between the particle
and the graphene, and describe the particle effectively using its electrical
polarizability which is expressed as $\alpha(\omega)=4\pi
R^3[\varepsilon(\omega)-1)]/[\varepsilon(\omega)+2]$~\cite{zhao2012rotational}. It is important to note that these
assumptions are valid only under the condition of $R\ll d \ll \hbar c / k_BT$.
In Fig.~\ref{fig2}(a), we present the imaginary part of the polarizability for these nanoparticles, revealing distinct dipolar resonances.

Now we study the effect of drift velocity on the net power transfer experienced by
different nanoparticles, as shown in
Fig.~\ref{fig2}(b). We differentiate between positive and negative energy fluxes, which
represent the flow of energy from the graphene to the particle and from the particle to
the graphene, respectively. Our analysis reveals intriguing behaviors. For NaCl and InSb
particles, the power transfer initially increases as $v_d$ increases. However, with increasing $v_d$ further, the power transfer decreases. In contrast, the direction of net power transfer undergoes a change for GaAs particles as the drift velocity increases. These are due to the interplay between the nonreciprocity of the SPPs and photon occupation number of graphene.

\begin{figure*}
\centering
\includegraphics[width=\textwidth]{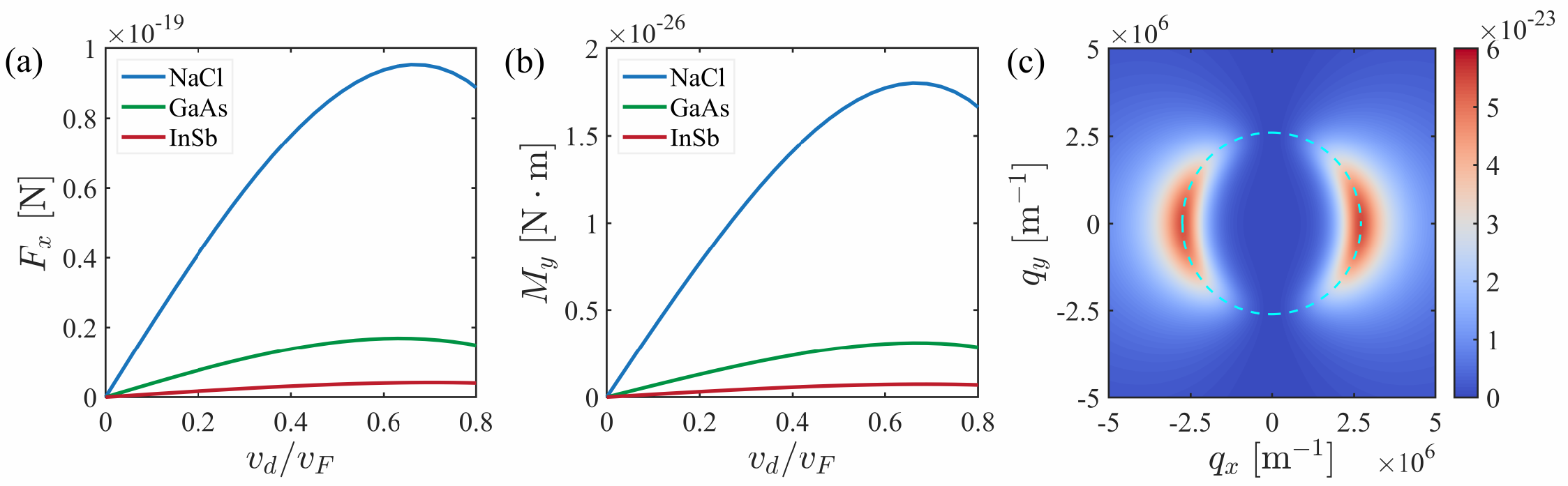} \\
\caption{(a) Net force on different nanoparticles as a function of $v_d$ at $d = 0.2\,\mu$m.
  (b) Net torque as a function of $v_d$ at $d = 0.2\,\mu$m.
  (c) Torque spectral density $m_y$ (in units of nJ$\,$nm$^2\,$s) as a function of $q_x$ and $q_y$ at $\hbar\omega =27.8\,$meV and $v_d=0.3v_F$. The cyan dashed line represents the dispersion of graphene SPPs.}
\label{fig3}
\end{figure*}

To analyze the contributions of forward ($q_x{>}0$) and backward ($q_x{<}0$) SPPs to net power
transfer, we use the NaCl particle as an example. In Fig.~\ref{fig2}(c), we show the
energy transmission function and the dispersion of graphene plasmons (shown by the cyan dashed line). This visual representation aligns the bright branches of the energy
transmission function with the cyan dashed line, confirming the dominant role played by
SPPs in governing power transfer. For $q_x{>}0$, the energy transfer
coefficients are positive, indicating energy transfer from graphene to the particle. For
$q_x{<}0$, the energy transfer coefficients are negative, signifying energy transfer from
the particle back to the graphene. The net energy transfer is the sum of
contributions from $q_x{>}0$ and $q_x{<}0$.  At lower drift velocities, the nonreciprocal
characteristics of SPPs are not pronounced, and energy
transfer is mainly governed by the nonreciprocal photon occupation number difference
$\delta n(\omega, q_x)$, as shown in Fig.~\ref{fig1}(c). However, as the drift velocity
increases, nonreciprocal SPPs interplay with $\delta n(\omega, q_x)$. The high drift velocities suppress the net power transfer due to the nonreciprocal photon occupation number even change the energy transfer direction between GaAs and graphene.

In the following, we investigate the transfer of linear and angular momenta
between the nanoparticles and the current-biased graphene. We define positive $M_y$  along the negative $y$ direction and
positive $F_x$ as oriented in the positive $x$ direction. Our findings revealed a
nonmonotonic behavior: Both force [Fig.~\ref{fig3}(a)] and torque [Fig.~\ref{fig3}(b)] increase with increasing the drift velocity up to a certain value, then start to decrease.
This behavior is consistent with the energy transfer.

\begin{figure}
\centering
\includegraphics[width=\columnwidth]{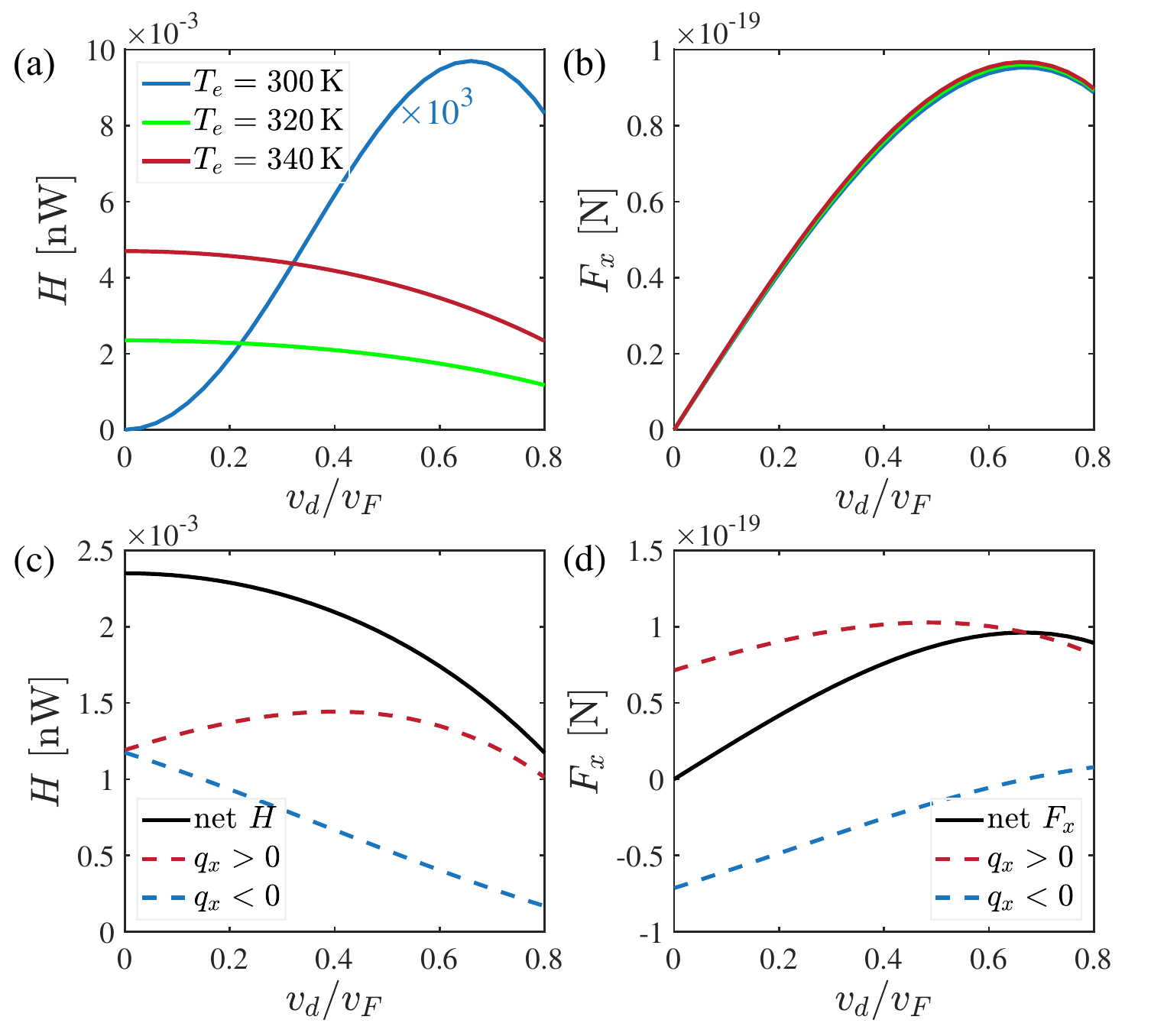} \\
  \caption{(a) Net power and (b) net force for the NaCl particle as a function of $v_d$ at various
  temperatures $T_e$ in graphene with $T_p=300\,$K and $d=0.2\,\mu$m. The curve of power transfer at $T_e=300\,$K is multiplied by $10^{3}$ for better visibility.
  (c) Net power and (d) net force vs $v_d$ with $T_e=320\,$K and $T_p=300\,$K. The contributions from the forward ($q_x{>}0$) and backward ($q_x{<}0$) SPPs are separately shown.  }
\label{fig4}
\end{figure}

To have a deeper understanding of the force and torque acting on the particle, we
analyze the exchange of linear momentum and spin angular momentum between the forward
($q_x{>}0$) and backward ($q_x{<}0$) SPPs and the nanoparticle. When photons transfer from the
graphene to the particle ($q_x{>}0$), the particle gains momentum, contributing to the
lateral force along the positive $x$ direction. Conversely, when $q_x{<}0$, the
particle loses momentum due to the reverse photon transfer, producing a lateral force
in the same direction. Therefore, SPPs with both positive and negative $q_x$
generate a net lateral force directed along the positive $x$ direction. Due to the
spin-momentum locking of SPPs~\cite{bliokh2015quantum}, the transverse spin of SPPs strongly
couples with their propagation direction. As a result, when SPPs propagate in the positive
(negative) $x$ direction, their spin aligns with the negative (positive) $y$ direction,
respectively. This spin angular momentum transfer is facilitated by photon transfer from
the graphene to the particle in the case of $q_x{>}0$, generating a torque directed along
the negative $y$ direction. Importantly, the torque contributed from $q_x{<}0$ is also along the negative $y$ direction. Consequently, the net lateral torque is along the
negative $y$ direction. These are supported by
Fig.~\ref{fig3}(c) which shows the torque spectral density $m_y$ as a function of $q_x$
and $q_y$ at $\hbar\omega = 27.8\,$meV and $v_d=0.3v_F$. Notably, for both
positive and negative $q_x$, the torque spectral densities $m_y$ are
positive.
Remarkably, the force and torque induced by current-biased graphene are of the same order of magnitude as those generated by thermally excited nonreciprocal surface electromagnetic waves~\cite{lateral_21-2}.

Finally, we consider the scenario where the temperature of the particle differs from that of
the graphene. Figures~\ref{fig4}(a) and ~\ref{fig4}(b) show the dependence of net energy transfer and lateral
force on the NaCl particle at various temperatures of the
graphene. The nanoparticle temperature is at $300\,$K. Notably, the
energy transfer from the graphene to the nanoparticle increases by three orders of
magnitude. To explain this, we calculate the contributions from both the forward and backward SPPs at $T_e{>}T_p$ [see Figs.~\ref{fig4}(c) and (d)].
It can be seen from Fig.~\ref{fig4}(c) that the energy transfer for photons with opposite momentum has the same sign, which is completely different from the case in the absence of a temperature difference. This means that the contribution from the temperature difference dominates over that from the electric current.
Therefore, the energy transfer is greatly enhanced in the presence of a temperature difference.
In contrast, the photons with $q_x{>}0$ and $q_x{<}0$ make positive and negative
contributions to the force, respectively. We find that the force experienced by the particle does not differ significantly from that without a temperature difference.
The torque has similar behavior as the force.
Thus, the temperature difference has a marginal impact on the lateral
force and torque while significantly enhancing the net power transfer with $T_e{>}T_p$.

\section{Conclusion}
We have investigated the near-field radiative energy, linear-momentum, and angular-momentum transfer from a current-biased graphene sheet to nanoparticles.
Notably, for the GaAs particle, both the magnitude and the direction of energy transfer can be electrically manipulated. This unique phenomenon
results from the interplay of nonreciprocal photon occupation differences and graphene surface plasmon polaritons.
Moreover, in the presence of a temperature difference, our study reveals distinct energy transfer
behavior while maintaining consistent force characteristics compared to the scenario without a temperature difference.
Recently, measuring energy transfer, force, and torque
between nanoparticles and surfaces has been experimentally realized~\cite{Menges_2016, Peng_2023, Jonghoon_2020}, so our theoretical predictions can thus be realized in the near future.

\begin{acknowledgments}
H.Z., L.Z., and J.C. acknowledge the support from the National Natural Science Foundation of China (Grants No.
12074230, No. 12174231), the Fund for Shanxi “1331  Project,” Fundamental Research Program of Shanxi Province
through 202103021222001, Program of Education and Teaching Reform in Shanxi Province (Grant No. J20230003), and
Research Project Supported by Shanxi Scholarship Council of China. G.T. is supported by National Natural Science Foun
dation of China (Grants No. 12374048 and No. 12088101) and NSAF (Grant No. U2330401). This research was partially
conducted using the High Performance Computer of Shanxi University.
\end{acknowledgments}

\bigskip
\noindent{$^{*}$gmtang@gscaep.ac.cn}\\
\noindent{$^{\dagger}$zhanglei@sxu.edu.cn}\\
\noindent{$^{\ddag}$chenjun@sxu.edu.cn}

\appendix
\section{Derivation of the spectral densities of the power transfer, the force and the torque}
From the fluctuational electrodynamics, the spectral densities of the net power transfer from the particle to the environment $h$, the force $f_j$, and the torque $m_j$ can be expressed as~\cite{henkel2002radiation,manjavacas2017lateral}
\begin{align}
  h(\omega)&= \omega {\rm Im}\Big[\Big\langle p_{j}^{{\text{fl}}^*}(\omega) E_{j}^{\text{ind}}(\omega)+ p_{j}^{{\text{ind}}^*}(\omega) E_{j}^{{\text{fl}}}(\omega)\Big\rangle\Big],
\label{Eqh} \\
  f_j(\omega)&= {\rm Re}\Big[\Big\langle p_k^{\text{fl}}(\omega) \partial_j E_k^{\text{ind}^*}(\omega)+ p_k^{\text{ind}}(\omega) \partial_j E_k^{{\text{fl}}^*}(\omega)\Big\rangle\Big],
\label{Eqf} \\
  m_j(\omega) &= \epsilon_{jkl} {\rm Re}\Big[\Big\langle p_k^{{\text{fl}}}(\omega) E_l^{{\text{ind}}^*}(\omega)+ p_k^{\text{ind}}(\omega) E_l^{{\text{fl}}^*}(\omega)\Big\rangle\Big],
\label{Eqm}
\end{align}
where the Einstein summation convention is applied with $j$,$k$,$l$ $\in$\{$x$, $y$, $z$\}, and $\langle\cdots\rangle$ denotes the statistical ensemble average. $E^{\mathrm{fl}}$ is the electric field generated by current fluctuations in the surface, $p^{\mathrm{fl}}$ is the dipole moment generated by the fluctuation of the particle dipole moment,
 $E^{\mathrm{ind}}$ is the field induced by the particle dipole moment fluctuations, and $p^{\mathrm{ind}}$ is the dipole moment induced in the particle by the environment field fluctuations. We assume that the nanoparticle is located at $\bm{r}$, and the dependence of $\bm{r}$ for the quantities on the right-hand sides are not written out explicitly.

The dipole moment induced by the electric-field fluctuations is given by
\begin{equation} \label{pind}
p_j^{\text{ind}}(\bm{r}) = \epsilon_0 \alpha_{j k} E_k^{\mathrm{fl}}(\bm{r}) ,
\end{equation}
where $\alpha$ is the polarizability of the particle.
The induced field by the dipole fluctuations can be expressed as
\begin{equation} \label{Eind}
E_{j}^{\mathrm{ind}}(\mathbf{r}_2, \omega)=\omega^2 \mu_{0} G_{j k}(\mathbf{r}_2, \mathbf{r}, \omega) p_{k}^{\mathrm{fl}}(\mathbf{r}, \omega),
\end{equation}
where the Green's function in the dipole-plate geometry is expressed as~\cite{nonrecip_torque19}
\begin{equation}
G(\mathbf{r}_2,\mathbf{r}_1,\omega)= \int \frac{d^2 {\bm q}}{(2\pi)^2} e^{i{\bm q} \cdot \big(\mathbf{R}_2-\mathbf{R}_1\big)} G(\mathbf{q},z_1+z_2, \omega) ,
\end{equation}
with
\begin{equation}
G(\mathbf{q},z_1+z_2,\omega)=
e^{i \beta _0 (z_1 + z _2)} \frac{ir_p}{2\beta_0} \hat{\mathbf{e}}_{p+} \hat{\mathbf{e}}_{p-}^T .
\end{equation}
Here, we use the notation $\mathbf{r}_n = (\mathbf{R}_n,z_n)$ with $n=1,2$.
The polarization vectors $\hat{\mathbf{e}}^T_{p \pm}$ for the $p$ polarized mode along the $\pm z$ direction are
\begin{equation}
\hat{\mathbf{e}}^T_{p \pm}=\frac{1}{k_0}\left(
\mp \beta_0 \cos \phi, \
\mp \beta_0 \sin \phi, \
q \right) .
\end{equation}
The in-plane wave vector is ${\bm q}=(q_x, q_y) =  q (\cos\phi,\sin\phi)$ where $\phi$ is the angle with respect to the $x$-axis. The out-of-plane wave vector in air is $\beta_0 = \sqrt{k_0^2 -q^2}$ with $k_0=\omega/c$. The Fresnel reflection coefficient for $p$ polarization is denoted as $r_p$.

The correlations functions of the fluctuating quantities, $p^{\mathrm{fl}}$ and $E^{\mathrm{fl}}$, respectively, satisfy the following fluctuation-dissipation relations with~\cite{landau2013statistical}
\begin{align}
\Big\langle p_j^{\mathrm{fl}}(\mathbf{r},\omega) p_k^{\mathrm{fl}^*}(\mathbf{r}, \omega') \Big\rangle =
& (1/2i)\hbar \epsilon_0\big[\alpha_{j k}(\omega)-\alpha_{k j}^*(\omega) \big]  \times \notag \\
& \bigg[n(T_p,\omega)+\frac{1}{2} \bigg] 2\pi\delta(\omega-\omega'),
\end{align}
and
\begin{align}
&\Big\langle E_j^{\mathrm{fl}}(\mathbf{r}_1, \omega)
 E_k^{\mathrm{fl}^{*}}(\mathbf{r}_2, \omega')\Big\rangle = (1/2i)\int \frac{d^2 {\bm q}}{(2\pi)^2}
 \bigg[ e^{i{\bm q} \cdot \big(\mathbf{R}_1-\mathbf{R}_2\big)}
 \notag \\
 &  G_{j k}(\mathbf{q}, z_1+z_2, \omega)-e^{i{\bm q} \cdot \big(\mathbf{R}_2-\mathbf{R}_1\big)}
 G_{k j}^*(\mathbf{q}, z_1+z_2, \omega) \bigg] \times
\notag \\
& \Big(\hbar \mu_0 \omega^2 \Big)\bigg[n(T_e, \omega -q_x v_d)+\frac{1}{2} \bigg]
2\pi \delta(\omega-\omega')\delta(\mathbf{r}_2-\mathbf{r}_1),
\end{align}
where $n(T,\omega)= \left[\exp(\hbar\omega/{k_B T})-1 \right]^{-1}$ is the Bose-Einstein distribution function at temperature $T$. Here, the correlation functions are antisymmetrized. The temperatures of the nanoparticle and the graphene are, respectively, denoted as $T_p$ and $T_e$.
The photon occupation number radiated from the graphene is Doppler shifted due to the electric current induced  drift velocity $v_d$ in the $x$ direction~\cite{VP08_PRB,VP11_PRL,VP11_PRB, drift1, GT21}.
To get equal frequency correlations, the factor of $2\pi$ in front of the delta functions in the above correlation functions is canceled by integrating over frequency $\omega'$.

The radiation power spectral density can be obtained as
\begin{align}
h(\omega)
&=\omega {\rm Im}\Big[\omega^2 \mu_{0}\Big\langle p_j^{\text{fl}^*}(\mathbf{r}, \omega) p_k^{\text{fl}}(\mathbf{r}, \omega) \Big\rangle G_{j k}(\mathbf{r}, \mathbf{r}, \omega) \notag \\
&\quad \quad \quad +\epsilon_0 \alpha^*(\omega)\Big\langle E_j^{\text{fl}^*}(\mathbf{r}, \omega) E_j^{\text{fl}}(\mathbf{r}, \omega)\Big\rangle\Big]
\notag \\
=&-\frac{\hbar\omega^3}{c^2} {\rm Im}[\alpha(\omega)] \int \frac{d^2 {\bm q}}{(2\pi)^2} {\rm Im}\big[G_{j j}(\mathbf{q}, 2d, \omega)\big] \delta n(\omega,q_x) ,
\end{align}
with the photon-occupation difference
\begin{equation}
  \delta n(\omega, q_x) = [e^{\hbar(\omega - q_x v_d)/k_B T_e} -1]^{-1}- [e^{\hbar\omega/k_B T_p} -1]^{-1}.
\end{equation}
We define the energy transmission function ${\cal Z}_h (\omega, {\bm q})$ through
\begin{equation}
h(\omega)
= \int \frac{d^2 {\bm q}}{(2\pi)^2} {\cal Z}_h(\omega, {\bm q}),
\end{equation}
so that
\begin{equation}
 {\cal Z}_h \left(\omega, {\bm q}\right) = -\hbar\omega k_0^2 {\rm
  Im} \big[\alpha(\omega)\big]\Theta(\omega, {\bm q}) \delta n(\omega, q_x),
\end{equation}
with
\begin{equation}
  \Theta(\omega, {\bm q}) = {\rm Re}\Big[r_p e^{2i\beta_0 d} \big(2q^2 -k_0^2\big)/(2\beta_0 k_0^2) \Big] .
\end{equation}

The spectral density of force can be obtained as
\begin{align}
& f_x(\omega)
={\rm Re}\Big[\Big\langle p_k^{\text{fl}}(\mathbf{r},\omega) \partial_{x}\big(\omega^2 \mu_0 G_{k j}(\mathbf{r}, \mathbf{r}, \omega) p_j^{\text{fl}}(\mathbf{r},\omega)\big)^*
\notag \\
&\quad \quad +\epsilon_0 \alpha(\omega) E_k^{\text{fl}}\left(\mathbf{r}, \omega\right) \partial_x E_k^{\text{fl}^*}(\mathbf{r}, \omega)\Big\rangle \Big] \notag \\
=&\hbar k_0^2 {\rm Im}\big[\alpha(\omega)\big]
 \int \frac{d^2 {\bm q}}{(2\pi)^2} q_x {\rm Im}\big[G_{k k} (\mathbf{q}, 2d, \omega)\big] \delta n(\omega, q_x) .
\end{align}
In the above, $\partial_x$ represents the partial derivative in the $x$ direction of position $\mathbf{r}$.
The spectral density of torque can be obtained as
\begin{align}
m_y(\omega)
=&\omega^2 \mu_{0} {\rm Re}\Big[\Big\langle p_z^{\text{fl}}(\mathbf{r},\omega) G_{x z}^*(\mathbf{r},\mathbf{r},\omega) p_z^{\text{fl}^*}(\mathbf{r},\omega)  \Big\rangle  \notag \\
& -\Big\langle p_x^{\text{fl}}(\mathbf{r}, \omega) G_{z x}^*(\mathbf{r},\mathbf{r},\omega) P_x^{\text{fl}^*}(\mathbf{r},\omega) \Big\rangle \Big] \notag \\
& +\epsilon_0 {\rm Re}\Big[\alpha(\omega) \Big\langle E_z^{\text{fl}}(\mathbf{r},\omega) E_x^{\text{fl}^*}(\mathbf{r},\omega) \Big\rangle
\notag \\
& -\alpha(\omega)\Big\langle E_x^{\text{fl}}(\mathbf{r},\omega) E_z^{\text{fl}^*}(\mathbf{r},\omega) \Big\rangle\Big] \notag \\
=&\hbar k_0^2 {\rm Im}\big[\alpha(\omega)\big]
\int \frac{d^2 {\bm q}}{(2\pi)^2}
{\rm Re}\Big[G_{z x}(\mathbf{q}, 2d, \omega)
\notag \\
& - G_{x z}(\mathbf{q}, 2d, \omega)\Big] \delta n(\omega, q_x) .
\end{align}
We thus have
\begin{equation}
f_x(\omega) = \int \frac{d^2 {\bm q}}{(2\pi)^2} {\cal Z}_f(\omega, {\bm q}), \
m_y(\omega) = \int \frac{d^2 {\bm q}}{(2\pi)^2} {\cal Z}_m(\omega,{\bm q}),
\end{equation}
with the force and the torque transmission functions, respectively, as
\begin{align}
 & {\cal Z}_f (\omega, {\bm q}) = \hbar k_0^2 q_x {\rm
  Im}\big[\alpha(\omega)\big] \Theta(\omega, {\bm q}) \delta n(\omega, q_x) , \\
 & {\cal Z}_m(\omega, {\bm q}) = -\hbar q_x {\rm Im}\big[\alpha(\omega)\big] {\rm Im}\big(r_p e^{2i\beta_0 d}\big) \delta n(\omega, q_x) .
\end{align}

\bibliography{lateral_fm_Fizeau_drag}{}

\end{document}